\documentclass[nohyperref]{article}

\usepackage{microtype}
\usepackage{graphicx}
\usepackage{subfigure}
\usepackage{booktabs} 

\usepackage{hyperref}
\usepackage{capt-of}

\newcommand{\proposed}{\text{scFP}}
\usepackage{algorithmic}
\usepackage{algorithm}

\usepackage[accepted]{icml2023}

\usepackage{amsmath}
\usepackage{amssymb}
\usepackage{mathtools}
\usepackage{amsthm}

\usepackage[capitalize,noabbrev]{cleveref}

\theoremstyle{plain}

\theoremstyle{definition}

\theoremstyle{remark}

\usepackage[textsize=tiny]{todonotes}

\begin{document}

\twocolumn[
\icmltitle{Single-cell RNA-seq data imputation using Feature Propagation}


\icmlsetsymbol{equal}{*}

\begin{icmlauthorlist}
\icmlauthor{Sukwon Yun}{equal,KAIST}
\icmlauthor{Junseok Lee}{equal,KAIST}
\icmlauthor{Chanyoung Park}{KAIST}
\end{icmlauthorlist}

\icmlaffiliation{KAIST}{KAIST}


\icmlcorrespondingauthor{Chanyoung Park}{cy.park@kaist.ac.kr}

\icmlkeywords{Machine Learning, ICML}

\vskip 0.3in
]



\printAffiliationsAndNotice{\icmlEqualContribution} 

\begin{abstract}
While single-cell RNA sequencing provides an understanding of the transcriptome of individual cells, its high sparsity, often termed dropout, hampers the capture of significant cell-cell relationships. Here, we propose \proposed~(single-cell Feature Propagation), which directly propagates features, i.e., gene expression, especially in raw feature space, via cell-cell graph. Specifically, it first obtains a warmed-up cell-gene matrix via Hard Feature Propagation which fully utilizes known gene transcripts. Then, we refine the $k$-Nearest Neighbor ($k$NN) of the cell-cell graph with a warmed-up cell-gene matrix, followed by Soft Feature Propagation which now allows known gene transcripts to be further denoised through their neighbors. Through extensive experiments on imputation with cell clustering tasks, we demonstrate our proposed model,~\proposed, outperforms various recent imputation and clustering methods. The source code of~\proposed~can be found at \url{https://github.com/Junseok0207/scFP}.
\end{abstract}

\section{Introduction}
Single-cell RNA-sequencing (scRNA-seq) analysis has attracted significant attention due to its property to profile transcriptome-wide gene expression at single-cell resolution. It allows researchers to perform various analyses, including identifying cell types~\cite{scdeepcluster,scgpcl}, and inferring cell trajectories~\cite{monocle, sctep}. However, analyzing scRNA-seq data is challenging due to the noisy nature of the gene expression. Specifically, scRNA-seq data often suffer from low transcript capture, referred to as dropout phenomena~\cite{dropout}, which causes the occurrence of false zero values. Furthermore, the observed non-zero expression values also suffer from biologically irrelevant signals, such as batch effects~\cite{batcheffect} and transcriptional noise~\cite{noise}.

Many works have been proposed to denoise the scRNA-seq data by imputing the dropout values. Among them, smoothing-based methods assume cells with similar expression profiles will likely share similar underlying biological characteristics. By leveraging this assumption, these methods impute gene expression values by forcing the expression values of neighboring cells to be more similar. Specifically, DrImpute~\cite{drimpute} averages the expression values based on a pre-calculated cluster, and MAGIC~\cite{magic}~performs a diffusing process on the calculated Markov affinity-based graph. Despite its effectiveness in reducing noise from various factors, there are some limitations to be considered: 1) it can decrease the meaningful cell variability by smoothing expression values across incorrect cell groups when neighboring cells are misdefined, and 2) noise expression values could be spread during smoothing when the observed expression values are noisy, which can be severe if it contains many false zero values due to the dropout phenomena.

Recently, most researchers are focused on the imputation methods that utilize deep neural networks (DNNs) to reconstruct gene expressions. These lines of methods utilize an autoencoder architecture, where the cell representation is learned through an encoder, and then impute values by passing them through the decoder layer. Specifically, DCA~\cite{dca} reconstructs the gene expression by assuming zero-inflated negative binomial (ZINB) distribution and AutoClass~\cite{autoclass} further learns the classifier by providing pseudo-labels generated by pre-clustering. Furthermore, scGCL~\cite{scgcl}, inspired by AFGRL~\cite{afgrl}, leverages relationship information between cells using graph neural networks (GNNs) on a cell-cell graph and learns cell representations in a self-supervised manner. However, despite their complex modeling, the output of these methods often shows poor performance on the subsequent downstream tasks compared to that of raw expression values. We argue that this is because effectively optimizing complex DNN models on a given dataset requires appropriate hyper-parameter choices, which can be challenging in the context of imputation tasks due to the unsupervised nature.

In this paper, we propose~\proposed, a simple yet effective imputation method for scRNA-seq data with a bi-level feature propagation scheme. Specifically, we first impute the zero values using the feature propagation on the initially defined $k$NN graph while preserving the non-zero values that contain relatively lower noise to prevent the contamination of biological signals from prevalent false zero values. Using this warmed-up data, the $k$NN graph is further refined to capture the correct neighbors of each cell. We then apply another feature propagation module to denoise the bias or noise on non-zero values. Through experiments on both real and simulated scRNA-seq datasets, we demonstrate the effectiveness of~\proposed.

\begin{figure*}[!t]
	\centering
	\includegraphics[width=1\textwidth]{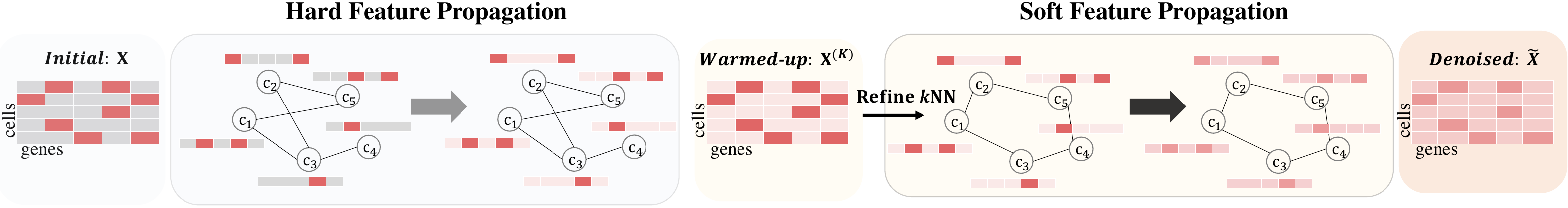}
        \vspace{-5ex}
        \caption{The overall architecture of \proposed.}
        \label{fig:architecture}
        \vspace{-3ex}
\end{figure*}

\section{Methods}

\noindent{\textbf{Notation.}} Given a cell-gene feature matrix $\mathbf{X} \in \mathbb{R}^{N\times M}$, where $N$ and $M$ are the number of cells and genes, let $\mathcal{G}=(\mathcal{V},\mathcal{E})$ denote a cell-cell graph. $\mathcal{V}=\{v_1,...,v_N\}$ and $\mathcal{E} \subseteq \mathcal{V} \times \mathcal{V}$ are the set of nodes and edges, respectively. $\mathbf{A} \in \mathbb{R}^{N \times N}$ is the adjacency matrix with $\mathbf{A}_{ij}=1$ iff $(v_i, v_j) \in \mathcal{E}$ and $\mathbf{A}_{ij}=0$ otherwise. We denote $\tilde{\textbf{A}}=\mathbf{D}^{-1}\mathbf{A}$ as a row-stochastic adjacency matrix, and the graph Laplacian matrix as $\boldsymbol{\Delta}=\mathbf{I}-\tilde{\mathbf{A}}$, where $\mathbf{I}$ is the identity matrix.

\subsection{Preliminary: Feature Propagation}

Recently, Feature Propagation (FP) \cite{fp} has been introduced to mitigate missing features in the graph domain. The core idea of FP is to diffuse the features that we know, i.e., provided, to the features that we do not know, i.e., missing, while maintaining the initial state of known features. Formally, given a node feature matrix consisting of $k$ known features and $u$ unknown features in each feature dimension $f$, the gradient flow of Dirichlet energy, i.e., $\ell(\mathbf{X}_{\cdot,f},\mathcal{G})= \frac{1}{2}\mathbf{X}_{\cdot,f}^{\top}\boldsymbol{\Delta}\mathbf{X}_{\cdot,f} $, at time step $t$ can be expressed as the heat diffusion equation:
\vspace{-0.3ex}
\begin{equation}
\label{eqn:heat_diffusion}
\begin{gathered}
\dot{\mathbf{X}}(t)=-\boldsymbol\nabla\ell(\mathbf{X}(t))	=-\boldsymbol{\Delta} \mathbf{X}(t), \\
(\text{IC}) \mathbf{X}_{\cdot,f}(0)=\left[\begin{array}{c} \small \mathbf{X}_{k,f}\\ \mathbf{X}_{u,f}(0)\end{array}\right], 
(\text{BC}) \mathbf{X}_{k,f}(t)=\mathbf{X}_{k,f}, \\
\forall k \in \mathcal{V}_{k,f}, \forall u \in \mathcal{V}_{u,f}, \forall f \leq F
\end{gathered}
\end{equation}

\noindent where $\mathbf{X}_{\cdot,f}\in \mathbb{R}^N$ denotes feature vector at dimension $f$ bounded by $F$, (IC) and (BC) denotes initial and boundary conditions, respectively. $\mathcal{V}_{k,f}$ and $\mathcal{V}_{u,f}$ denotes a set of \textit{known} nodes and \textit{unknown} nodes at feature dimension, $f$, respectively. Here, solving Equation~\ref{eqn:heat_diffusion} by linear equation, we obtain closed-from solution, $\mathbf{X}_{u}=-\boldsymbol{\Delta}_{uu}^{-1}\boldsymbol{\Delta}_{ku}^{\top}\mathbf{X}_{k}$. However, during calculation, it induces complexity of $\mathcal{O}(|\mathcal{V}_{u}|^3)$, which is not desirable in large graphs. Thus, we resort to an iterative Euler scheme and derive a formula as follows:

\begin{equation}
\label{eqn:iterative}
\begin{gathered}
\begin{aligned}
\mathbf{X}^{(i+1)}  &=
\left[\begin{array}{cc}
\mathbf{I}  & \mathbf{0} \\
-\boldsymbol{\Delta}_{uk} & \mathbf{I}-\boldsymbol{\Delta}_{uu}
\end{array}\right] \mathbf{X}^{(i)} \\
&= \left[\begin{array}{cc}
\mathbf{I} & \mathbf{0} \\
\tilde{\mathbf{A}}_{uk} & \tilde{\mathbf{A}}_{uu}
\end{array}\right] \mathbf{X}^{(i)} 
\end{aligned}
\end{gathered}
\end{equation}


\noindent where $\mathbf{X}^{(i)}$ represents imputed feature matrix at $i$ step. It is important to note that this formula is basically equivalent to multiplying the feature matrix with normalized adjacency while maintaining known feature values as its initial state. It shows the robust performance in the graph domain even when the data contains more than 90\% missing features~\cite{fp}. However, it cannot be naively applied to scRNA-seq data due to its following inherent nature: 1) The information regarding which features are missing or noisy is not provided. 2) As the graph structure is not provided, it is crucial to construct a graph that connects biologically relevant cells.


\subsection{Proposed Methodology: \proposed}
Here, we propose the bi-level feature propagation method that extends FP in a manner that is well-suited to the scRNA-seq domain. Our overall architecture can be found in Figure~\ref{fig:architecture}. We first define a cell-cell graph using initial scRNA-seq data and perform a diffusion process using a Hard Feature Propagation scheme (Sec~\ref{sec:hard_fp}) that primarily pays attention to imputing zero values in the cell-gene count matrix while preserving non-zero values. After that, we refine the cell-cell graph by calculating $k$-nearest neighbors for each cell (Sec~\ref{sec:refine_knn}) using the previously smoothed outputs (i.e., warmed-up data). Then, we pass through Soft Feature Propagation (Sec~\ref{sec:soft_fp}) which allows the denoising of observed gene transcripts. Detail about each component can be found in the following sections.


\subsubsection{Hard Feature Propagation}
\label{sec:hard_fp}
We start with imputing zero values, i.e., dropouts in gene expression with the aid of similar cells. To do so, we first need to define an adjacency matrix which first brings us a challenge compared to the graph domain, where adjacency information is provided. Here, the intuitive and cheap way that facilitates message-passing between similar cells is to introduce a $k$NN graph by calculating cosine similarities. However, considering the sparsity of cell-gene matrix \cite{magic, misc}, which is a direct resource for building $k$NN graph, we argue that obtained $k$NN graph should merely serve as an initialized graph for warming-up sparse and noisy gene transcripts. In this regard, as our primal goal is to impute zero-valued gene expressions (indexed by $z$) via non-zero-valued gene expressions (indexed by $n$), we apply Hard Feature Propagation as follows:
\begin{equation}
\label{eqn:hard_fp}
\mathbf{X}^{(i+1)}=\left[\begin{array}{cc}
\mathbf{I} & \mathbf{0} \\
\tilde{\mathbf{A}}_{zn}^{\text{initial}} & \tilde{\mathbf{A}}_{zz}^{\text{initial}}
\end{array}\right] \mathbf{X}^{(i)}
\end{equation}
\noindent where $\mathbf{X}^{(i)} \in \mathbb{R}^{N\times M}$ represents imputed cell-gene feature matrix at step $i$. After that, we obtain a converged warm-up matrix, $\mathbf{X}^{(K)}$, which is now denser and smoothed via neighbors. 
Note that in this step, we employ a hard clamping strategy where the non-zero values, which correspond to the observed expression values, are retained at their original values.
This is because, in this step, more emphasis is placed on the imputation of zero-values, which are more prevalent in the cell-gene count matrix, compared to denoising non-zero transcript values. This is because non-zero values generally contain less noise than zero-values, i.e., dropouts.




\subsubsection{Refining Cell-Cell Graph}
\label{sec:refine_knn}
With a warmed-up cell-gene matrix, $\mathbf{X}^{(K)}$ originating from Hard Feature Propagation, we make use of this matrix to refine an initial cell-cell graph which was built when cell representation was sparse. More precisely, a warmed-up cell-gene matrix is used as input for $k$NN graph as follows:
\begin{equation}
\label{eqn:refine}
\tilde{\mathbf{A}}^{\text{refined}} = k\text{NN}(\mathbf{X}^{(K)})
\end{equation}
\noindent where $\tilde{\mathbf{A}}^{\text{refined}} \in \mathbb{R}^{N\times N}$ is refined normalized adjacency for a cell-cell graph which is built considering the denoised zero-values, which was not feasible to capture in the initial stage of $k$NN graph generation. In other words, this process could potentially reveal hidden or implicit graph structures that were not initially detectable due to sparse and noisy gene expression. We argue that this refined adjacency matrix may provide more accurate or robust graph representations, which could enhance subsequent analyses or learning tasks.

\subsubsection{Soft Feature Propagation}
\label{sec:soft_fp}
Now, equipped with warmed-up cell-gene matrix, $\mathbf{X}^{(K)}$ and refined adjacency matrix for cell-cell graph, $\tilde{\mathbf{A}}^{\text{refined}}$, we hereby run Soft Feature Propagation. Specifically, compared to Hard Feature Propagation which is based on hard clamping \cite{lp_main, lp_var3, lp_var4}, at this moment, we rather adopt soft clamping \cite{lp_var1, lp_var2}, which basically leaves room for updating the originally transcripted gene value of its neighbors endowed with implicit graph structures thanks to the warmed-up procedure. This aligns with our motivation to take into account noise from not only low transcript capture but also from the observed non-zero expression values that might possess biologically irrelevant signals, e.g., batch effects and transcriptional noise~\cite{batcheffect, noise}. Formally, Soft Feature Propagation is applied as follows:
\begin{equation}
\label{eqn:soft}
\mathbf{X}^{(K+j+1)} = \alpha\tilde{\mathbf{A}}^{\text{refined}}\mathbf{X}^{(K+j)}+(1-\alpha)\mathbf{X}^{(K)}
\end{equation}
\noindent where $\mathbf{X}^{(K+j)}$ is the updated cell-gene matrix at step $j$, $0<\alpha<1$ is the constant\footnote{As we aim to denoise known values, i.e., not to maintain its original state, we fixed $\alpha$ as 0.99 and used as constant.} that controls amount of information that $\mathbf{X}^{(K+j)}$ receives from its neighbors. After another $K$ iteration, we finally obtain a converged and denoised cell-gene matrix $\tilde{\mathbf{X}} = \mathbf{X}^{(2K)}$. This denoised matrix considers both the noise from zero-values (dropouts) and the noise from non-zero-values (transcriptional noise), and it will serve as the main resource for subsequent downstream tasks. Detailed algorithm for~\proposed~can be found in Appendix~\ref{implementation details}.


\begin{table*}[t]
\caption{Overall performance in imputation task measured by RMSE.}
\centering  
\label{table:imputation}
\begin{small}
\resizebox{0.7\textwidth}{!}{
\begin{tabular}{c|ccc|ccc|ccc|ccc|ccc}
\toprule
              & \multicolumn{3}{c|}{Baron Mouse}   & \multicolumn{3}{c|}{Mouse ES}      & \multicolumn{3}{c|}{Mouse Bladder} & \multicolumn{3}{c|}{Zeisel}        & \multicolumn{3}{c}{Baron Human}   \\ \cline{2-16} 
              & \multicolumn{3}{c|}{Dropout Rates} & \multicolumn{3}{c|}{Dropout Rates} & \multicolumn{3}{c|}{Dropout Rates} & \multicolumn{3}{c|}{Dropout Rates} & \multicolumn{3}{c}{Dropout Rates} \\ \cline{2-16} 
              & 20\%        & 40\%        & 80\%      & 20\%        & 40\%        & 80\%      & 20\%        & 40\%       & 80\%       & 20\%        & 40\%       & 80\%       & 20\%        & 40\%       & 80\%      \\ \hline
MAGIC         & 0.61       & 0.73       & 0.99     & 0.53       & 0.73       & 1.21     & 0.50       & 0.60      & 0.82      & 0.60       & 0.82      & 1.31      & 0.58       & 0.74      & 1.06     \\
DCA           & 0.42       & 0.43       & 0.49     & 0.35       & 0.35       & 0.36     & 0.37       & 0.38      & 0.41      & 0.39       & 0.42      & 0.44      & 0.41       & 0.43      & 0.47     \\
AutoClass     & 0.63       & 0.76       & 0.98     & 0.53       & 0.75       & 1.23     & 0.52       & 0.64      & 0.82      & 0.60       & 0.84      & 1.32      & 0.59       & 0.76      & 1.08     \\
scGCL         & 0.64       & 0.74       & 0.97     & 0.59       & 0.75       & 1.16     & 0.51       & 0.62      & 0.81      & 0.66       & 0.82      & 1.29      & 0.63       & 0.77      & 1.08     \\ \hline
scFP (Ours)          & \textbf{0.36}       & \textbf{0.37}       & \textbf{0.43}     & \textbf{0.32}       & \textbf{0.32}       & \textbf{0.36}    & \textbf{0.26}       & \textbf{0.26}      & \textbf{0.31}      & \textbf{0.39}       & \textbf{0.40}      & \textbf{0.44}      & \textbf{0.33}       & \textbf{0.34}      & \textbf{0.39}   \\
\bottomrule
\end{tabular}
}
\end{small}
\end{table*}


\begin{table*}[t]
\caption{Overall performance of cell clustering task measured by ARI, NMI, CA.}
\centering
\label{table:cell_clustering}
\begin{small}
\resizebox{0.7\textwidth}{!}{
\begin{tabular}{c|ccc|ccc|ccc|ccc|ccc}
\toprule
              & \multicolumn{3}{c|}{Baron Mouse} & \multicolumn{3}{c|}{Mouse ES} & \multicolumn{3}{c|}{Mouse Bladder} & \multicolumn{3}{c|}{Zeisel} & \multicolumn{3}{c}{Baron Human} \\ \cline{2-16} 
              & ARI       & NMI       & CA       & ARI      & NMI      & CA      & ARI        & NMI       & CA        & ARI     & NMI     & CA      & ARI       & NMI      & CA       \\ \hline
Raw           & 0.44      & 0.71      & 0.56     & 0.74     & 0.75     & 0.79    & 0.59       & 0.75      & 0.68      & 0.70    & 0.75    & 0.77    & 0.44      & 0.71     & 0.56     \\
MAGIC         & 0.42      & 0.72      & 0.57     & 0.80     & \textbf{0.85}     & 0.83    & 0.55       & 0.75      & 0.64      & 0.70    & 0.75    & 0.76    & 0.56      & 0.78     & 0.59     \\
DCA           & 0.46      & 0.69      & 0.59     & 0.76     & 0.78     & 0.81    & 0.39       & 0.59      & 0.54      & 0.67    & 0.72    & 0.75    & 0.53      & 0.74     & 0.55     \\
AutoClass     & 0.44      & 0.71      & 0.52     & 0.74     & 0.75     & 0.81    & 0.51       & 0.75      & 0.64      & 0.71    & 0.75    & 0.77    & 0.44      & 0.71     & 0.52     \\
scGCL         & 0.43      & 0.72      & 0.54     & 0.73     & 0.75     & 0.79    & 0.53       & 0.75      & 0.64      & 0.65    & 0.70    & 0.73    & 0.50      & 0.78     & 0.62     \\
kNN-smoothing & 0.43      & 0.72      & 0.55     & 0.72     & 0.74     & 0.79    & 0.59       & 0.76      & 0.68      & 0.68    & 0.73    & 0.76    & 0.56      & 0.78     & 0.56     \\ \hline
scFP (Ours)          & \textbf{0.61}      & \textbf{0.82}      & \textbf{0.76}     & \textbf{0.82}     & 0.83     & \textbf{0.85}    & \textbf{0.65}       & \textbf{0.77}      & \textbf{0.73}      & \textbf{0.85}    & \textbf{0.81}    & \textbf{0.89}    & \textbf{0.68}     & \textbf{0.83}     & \textbf{0.73}    \\
\bottomrule
\end{tabular}
}
\end{small}
\end{table*}

\section{Experiments}

\noindent\textbf{Experimental Settings. } To evaluate the effectiveness of~\proposed, we evaluated~\proposed~with 5 widely used real-world scRNA-seq data, Baron Mouse, Mouse ES, Mouse Bladder, Zeisel, and Baron Human. The detailed statistics of these datasets can be found in Appendix~\ref{appendix:statistics}.

\noindent\textbf{Performance on Imputation. } Table~\ref{table:imputation} shows the overall performance in imputation tasks with dropout rates ranging from the low case, e.g., 20\%, to the severe case, e.g., 80\%. We observe~\proposed~shows robust performance regardless of dropout rates outperforming other baselines designed for denoising cell-gene matrix. Among baselines, it is worth noting that more complex models, e.g., AutoClass, and scGCL, exhibit lower performance than DCA. This tells us that the complexity of the model does not always guarantee imputation performance in the scRNA-seq domain. In this regard, the proposed method,~\proposed, does not require any learnable parameters and demonstrates its effectiveness by simply propagating features on the raw feature space. This highlights the importance of carefully handling observed values in order to obtain a well-imputed cell-gene matrix.



\noindent\textbf{Performance on Cell Clustering. } With obtained denoised cell-gene matrix, we further evaluate whether the denoised matrix performs well on the representative downstream task, cell clustering. As shown in Table~\ref{table:cell_clustering}, we observe~\proposed~achieve promising results in terms of Adjusted Rand Index (ARI), Normalized Mutual Information (NMI), and Clustering Accuracy (CA). Interestingly, while DCA performed well on imputation tasks compared to other baselines, its robustness could not be maintained in clustering tasks. It is also worth noting that \proposed~consistently outperforms the performance of MAGIC, demonstrating that the bi-level feature propagation and structure refinement are important to effectively denoise the scRNA-seq data. Furthermore,~\proposed~demonstrates its robustness by consistently exhibiting superior performance compared to raw data across all five datasets.

\begin{figure}[h]
	\centering
	\includegraphics[width=0.45\textwidth]{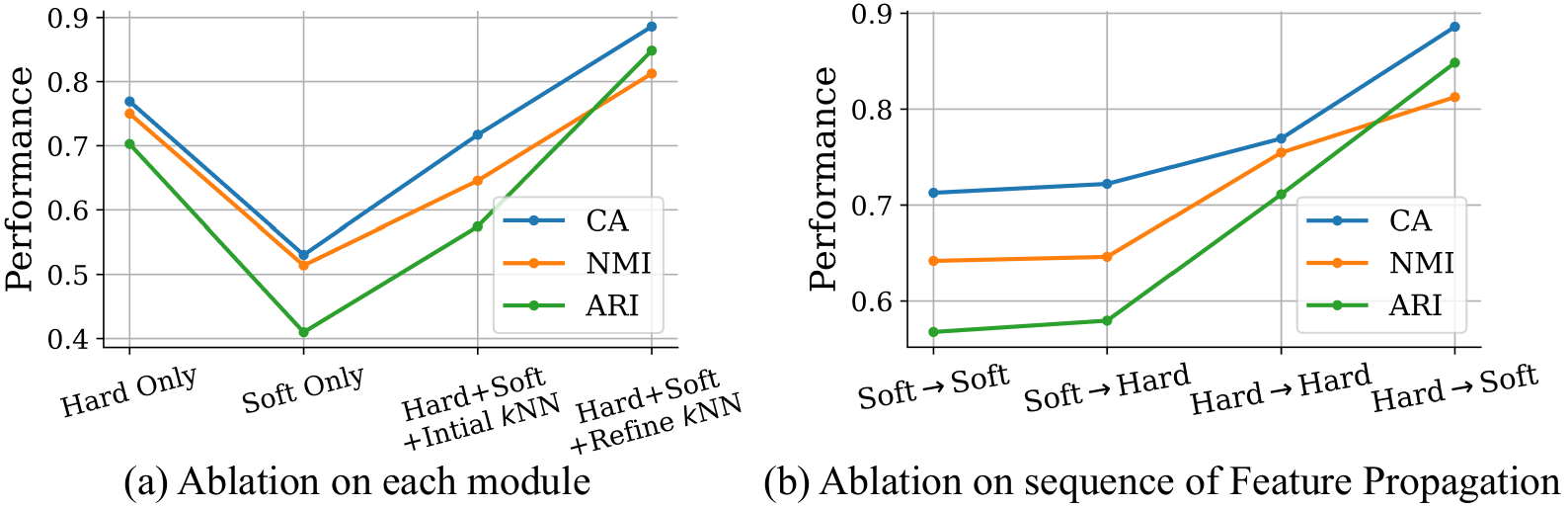}
        \vspace{-2ex}
        \caption{Ablation studies of~\proposed. (a) "Hard+Soft+Refine $k$NN" corresponds to~\proposed. (b) "Hard$\rightarrow$Soft" corresponds to~\proposed. Zeisel dataset is used.}
        \label{fig:ablation}
        \vspace{-1ex}
\end{figure}

\noindent\textbf{Ablation studies. } Figure~\ref{fig:ablation} shows the ablation studies from two perspectives. First, by incremental ablation on each module, we observe~\proposed~fully benefits when feature propagation has been made in both Hard and Soft ways. Here, it is important to note that utilizing a refined $k$NN graph, where a graph is reconstructed by a warmed-up matrix, is crucial before processing Soft Feature Propagation. Also, with ablation on the sequence of Feature Propagation, we verify the usage of Hard Feature Propagation, which maintains the observed value with its initial state at the early stage of imputation is significant. However, solely resorting to Hard Feature Propagation outputs sub-optimal results since it does not leave room for the observed values, which also possess noise, to be denoised via their neighbors.

\begin{figure}[t]
	\centering
	\includegraphics[width=0.35\textwidth]{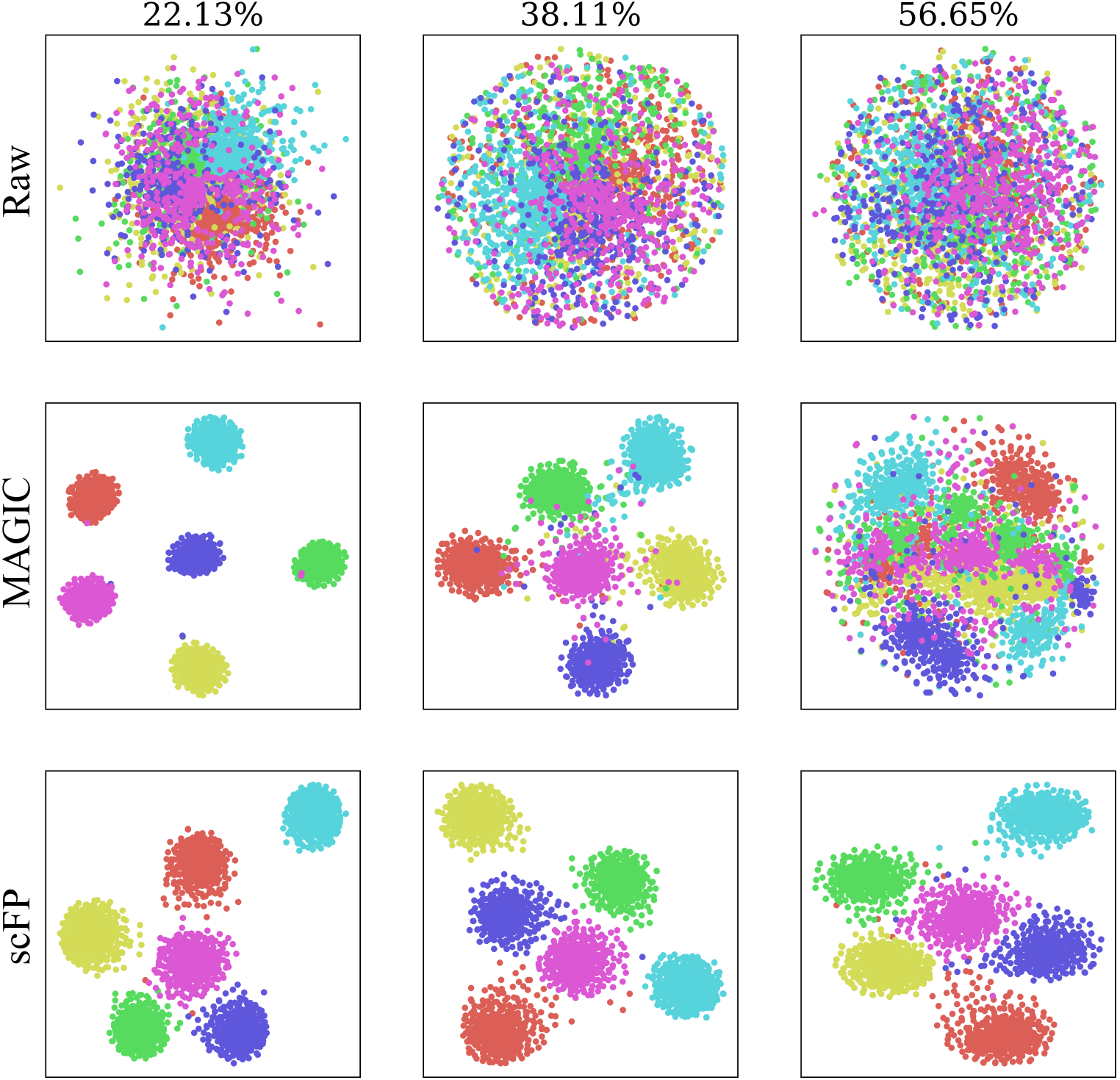}
         \vspace{-2.5ex}
        \caption{t-SNE visualization result on simulated dataset over various dropout rates. The rates at the top represent the dropout rate, which equals to the proportion of false zero values.}
        \vspace{-3.45ex}
        \label{fig:tsne}
\end{figure}

\noindent\textbf{Simulation studies. }
To demonstrate our claim that the diffusion of noise from false zeros can have a negative effect, we conducted experiments using a simulation dataset generated by Splatter Package~\cite{splatter}, where we can control the dropout rate, indicating the proportion of false zeros. In Figure~\ref{fig:tsne}, MAGIC, which diffuses both zero and non-zero values, successfully separates cell types by preserving biologically relevant signals when the dropout rate is relatively low (i.e., 22.13\%). However, in cases with a high dropout rate (i.e., 56.65\%), where a significant number of false zeros are present, it fails to separate cell types due to the contamination from false zeros. On the other hand,~\proposed~effectively separates cell types even in situations with a high dropout rate, thanks to the careful diffusion process of the Hard FP step, which preserves non-zero values. 

\section{Conclusion}
In this paper, we proposed~\proposed~which imputes and denoises the observed scRNA-seq data that is inherently sparse and noisy. Specifically, we first aimed to impute zero-values of transcripts in Hard Feature Propagation with hard clamping of observed values. With a warmed-up matrix, we then refined the $k$NN graph and proceeded with Soft Feature Propagation in order to denoise known values with its neighbors, taking into account the potential of transcriptional noise. With a simple and lightweight design, its imputation and cell clustering performance under various datasets verifies the effectiveness of~\proposed. 

\clearpage
\bibliography{scFP_reference}
\bibliographystyle{icml2023}

\newpage
\appendix
\onecolumn

\section{Data statistics}
\label{appendix:statistics}

\label{sec:data_statistics}
\begin{table}[h]
\centering
\caption{Statistics for real datasets used for experiments.}
\resizebox{0.6\linewidth}{!}{
\begin{tabular}{c|ccccc}
\noalign{\smallskip}\noalign{\smallskip}
\toprule
Data & \# of Cells & \# of Genes & \# of Subgroups \\
\hline
Baron Mouse & 1,886 & 14,861 & 13 \\
Mouse ES cells & 2,717 & 24,047 & 4 \\
Mouse Bladder cells & 2,746 & 19,771 & 16 \\
Zeisel & 3,005 & 19,972 & 7 \\
Baron Human & 8,569 & 20,125 & 14 \\
Shekhar Mouse Retina cells & 27,499 & 13,166 & 19 \\
\bottomrule
\end{tabular}
}
\label{tab:real dataset}
\end{table}

\section{Extension to large dataset}

\begin{table*}[h]
\begin{minipage}{0.45\linewidth}{
    \centering
    \small
    \caption{Performance of cell clustering in Shekhar Mouse Retina dataset.}
    \resizebox{0.9\linewidth}{!}{
        \begin{tabular}{c|ccc}
        \bottomrule
                      & \multicolumn{3}{c}{Shekhar Mouse Retina} \\ \cline{2-4} 
                      & ARI     & NMI     & CA      \\ \hline
        Raw           & 0.54    & 0.76    & 0.61    \\
        MAGIC         & 0.64    & 0.82    & 0.73    \\
        DCA           & 0.34    & 0.37    & 0.42    \\
        AutoClass     & 0.74    & 0.75    & 0.81    \\
        scGCL         & 0.54    & 0.74    & 0.58    \\
        kNN-smoothing & 0.47    & 0.75    & 0.55    \\ \hline
        scFP (Ours)         & \textbf{0.91}    & \textbf{0.83}    & \textbf{0.82}  \\
        \bottomrule
        \end{tabular}
        }
    \label{tab:bi_table}
} \end{minipage}
\begin{minipage}{0.5\linewidth}{
\centering
    \includegraphics[width=0.8\linewidth]{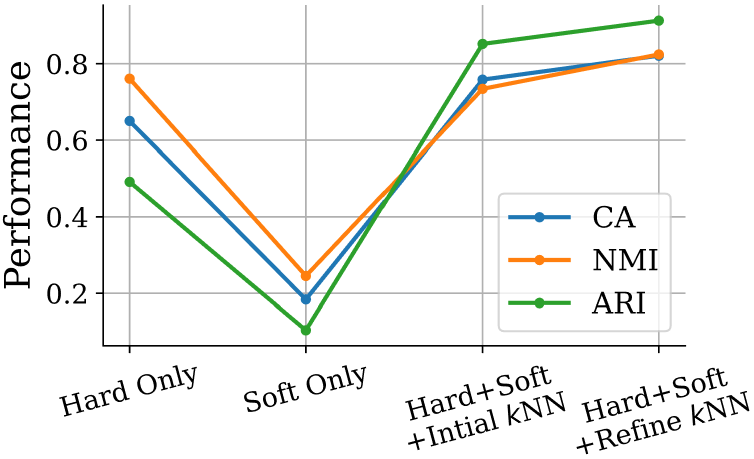} 
    \vspace{-2ex}
    \captionof{figure}{Ablation of each module in~\proposed~in Shekhar Mouse Retina dataset. "$\text{Hard}\text{+}\text{Soft}\text{+}\text{Refine $k$NN}$" denotes~\proposed.}
    \label{fig:bi_figure}
}\end{minipage}
\end{table*}

Here, we further extended our experiment on a relatively large dataset, Shekhar mouse retina cells, and compared the performance of cell clustering in Table~\ref{tab:bi_table} with its ablation study (Figure~\ref{fig:bi_figure}) on each module on~\proposed. Despite the absence of trainable parameters, our proposed method still demonstrates promising clustering performance on a large dataset. This suggests that, sometimes, rather than focusing on the complexity of the model in a trainable sense, giving careful consideration to the raw feature space and leveraging given resources, e.g., observed transcripts can be crucial.

\section{Pseudocode of the proposed method}
\label{implementation details}
\begin{algorithm}
\small
\caption{single-cell Feature Propagation (scFP)}\label{alg:scFP}
\begin{algorithmic}[1]

\STATE {\bfseries Input:} Cell-Gene Matrix $\mathbf{X}$, Initial $k$NN $\tilde{\mathbf{A}}^{\text{initial}}$\\
\STATE {\bfseries Output:} Denoised Cell-Gene Matrix $\tilde{\mathbf{X}}$ \\
\STATE $\mathbf{Y} \gets \mathbf{X}$ \\
\STATE \algorithmicwhile $ \; \mathbf{ X} \text{ has not converged}$ \algorithmicdo \\
\STATE \;\;\;\; $\mathbf{X} \gets \tilde{\mathbf{A}}^{\text{initial}}\mathbf{X}$ \\ 
\STATE \;\;\;\; $\mathbf{X}_{k,d} \gets \mathbf{Y}_{k,d} \forall k \in \mathcal{V}_{k,d}, \forall d \leq M$ \hfill $\triangleright$ Hard Clamping \\

\STATE \algorithmicendwhile

\STATE $\tilde{\mathbf{A}}^{\text{refined}} = k\text{NN}(\mathbf{X}^{(K)})$ \hfill  $\triangleright$ Refine $k$NN \\

\STATE \algorithmicwhile $\; \mathbf{X}^{(K)} \text{ has not converged}$ \algorithmicdo \\

\STATE \;\;\;\; $\mathbf{X}^{(K)} \gets \alpha\tilde{\mathbf{A}}^{\text{refined}}\mathbf{X}^{(K)}+(1-\alpha)\mathbf{X}^{(K)}$ \hfill  $\triangleright$ Soft Clamping

\STATE \algorithmicendwhile

\end{algorithmic}
\end{algorithm}

Observing that Equation~\ref{eqn:hard_fp} in Hard Feature Propagation essentially propagates features (i.e., gene expressions) via neighbors and resets the originally expressed gene values, we formulate the whole process in Algorithm~\ref{alg:scFP}. For the iteration until convergence, we used 40, which is enough to converge, as mentioned in FP \cite{fp}. It is important to note that we did not use any trainable parameters during the whole process and obtained a denoised matrix solely by raw feature space. Overall, in this work, we aim to emphasize simple and straightforward ways to enhance performance in imputation on subsequent downstream tasks, e.g., cell clustering.


\end{document}